\documentclass[conference]{IEEEtran}
\IEEEoverridecommandlockouts

\usepackage{cite}
\usepackage{algorithmic}
\usepackage{algorithm}
\usepackage{graphicx}
\usepackage{textcomp}
\usepackage{xcolor}
\usepackage{amsmath,amssymb,amsfonts}
\usepackage{url}

\def\BibTeX{{\rm B\kern-.05em{\sc i\kern-.025em b}\kern-.08em
    T\kern-.1667em\lower.7ex\hbox{E}\kern-.125emX}}
\begin{document}

\title{A Multi-Agent Reinforcement Learning Scheme for SFC Placement in Edge Computing Networks}

\author{\IEEEauthorblockN{Congzhou Li, Zhouxiang Wu, Divya Khanure, and Jason P. Jue}
    \IEEEauthorblockA{
        Department of Computer Science, The University of Texas at Dallas, Richardson, Texas 75080, USA\\        
    }
}

\maketitle

\begin{abstract}
In the 5G era and beyond, it is favorable to deploy latency-sensitive and reliability-aware services on edge computing networks in which the computing and network resources are more limited compared to cloud and core networks but can respond more promptly. These services can be composed as Service Function Chains (SFCs) which consist of a sequence of ordered Virtual Network Functions (VNFs). To achieve efficient edge resources allocation for SFC requests and optimal profit for edge service providers, we formulate the SFC placement problem in an edge environment and propose a multi-agent Reinforcement Learning (RL) scheme to address the problem. The proposed scheme employs a set of RL agents to collaboratively make SFC placement decisions, such as path selection, VNF configuration, and VNF deployment. Simulation results show our model can improve the profit of edge service providers by 12\% compared with a heuristic solution.
\end{abstract}


\section{Introduction}

Many emerging applications in 5G and beyond require the support of various combinations of network functions. Physical network functions may be virtualized as Virtual Network Functions (VNFs) and deployed in the cloud/edge computing networks. To complete specific tasks, the VNFs need to be chained together to form a Service Function Chain (SFC). For applications which require service with adequate reliability and low latency, deploying SFCs and serving them at the edge of the network can achieve task offloading and improved quality of service \cite{8016573}. When SFC requests arrive at the Access Point (AP) of an edge computing network, due to the stringent performance bound, the eligible paths for SFC placement is restricted to the local area of the AP with limited computing and network resources availability. For edge computing service providers, how to determine the placement plan for SFC requests to achieve optimal profit, while meeting the performance requirement of requests is an issue of significant importance. 

\begin{figure}[tb]
\centerline{\includegraphics[width=0.35\textwidth]{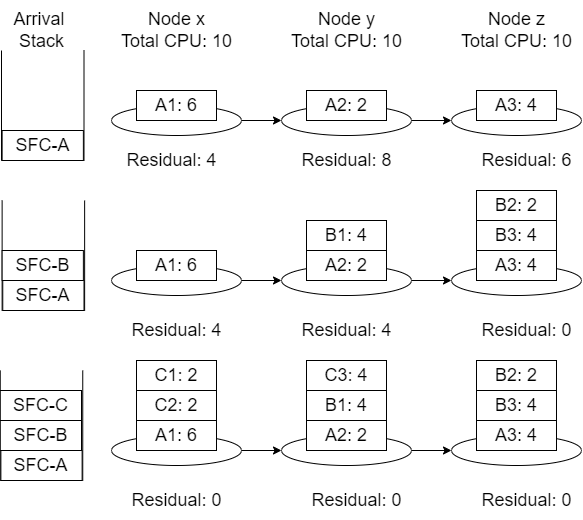}}
\caption{Example Deployment Patterns of Three SFCs }
\label{fig:pattern_sample}
\end{figure}

The end-to-end delay of a SFC comprises the transmission delay of path links and the processing delay of VNFs. In \cite{alleg2017delay}, the authors show that if the VNFs of a SFC are allocated additional computing resources, then increased transmission delays of this SFC can be offset by reduced processing delays of VNFs. Thus, in edge computing networks, for those SFC deployment plans whose total delay exceed the latency requirement, we can increase computing resources to VNFs to reduce their processing delays to make such plans become feasible. This delay compensation approach can provide more candidate paths for SFC requests and the computing resources allocated to VNFs will be path-specific.

Providing replicas for a VNF can increase its reliability. However, if each replica consumes the same amount of computing resources as the original VNF, the cost of reliable SFC deployment can be unreasonably expensive in an edge computing environment. In \cite{akahoshi2021service}, the VNFs can adjust their computing resource allocation during runtime. The replicas of a VNF can be allocated with minimum computing resource to remain alive, and can be activated with more resources if the original VNF fails. Thus, the reliability of a VNF can be improved by adding replicas with minimal resource consumption. 

In the SFC placement process which adopts delay compensation and VNF replicas, the network operator needs to form a routing path for the SFC request, configures computing resources allocated to each VNF and determines the deployment locations of VNFs along the path. The placement locations of VNFs on the path have various choices and we define each possible VNF placement combination as a deployment pattern of the SFC. A reasonable selection from possible deployment patterns can achieve efficient utilization of edge computing resources. For example, in Fig. \ref{fig:pattern_sample}, three SFC requests arrive at the edge computing network in time sequence and are deployed along the same path. By arranging the VNF placement patterns of ``SFC-B" and ``SFC-C", the CPU cores installed on each computing node are fully utilized, and all the SFC requests are accepted and deployed successfully.

In this paper, we propose a RL-based SFC placement scheme in which a group of agents collaborate to make admission, allocation and mapping decisions for the placement of SFC requests in order to maximize the profit of edge service providers. To the best of our knowledge, this paper is the first to apply a multi-agent RL-based scheme for the SFC placement problem while considering latency and reliability constraints in edge computing networks. The remainder of this paper is organized as follows. Section \ref{sec:Related_Works} introduces related works on SFC mapping problem. We describe the system model and formulate the problem in Section \ref{sec:problem_formulation}. Next, we illustrate our RL-based SFC placement scheme in Section \ref{sec:Proposed_Scheme}. In Section \ref{sec:experiment}, we design experiments to evaluate the effectiveness of the proposed scheme. Finally, we conclude the paper in Section \ref{sec:conclusion}.

\section{Related Works}
\label{sec:Related_Works}

In \cite{9236981}, the authors proposed a Q-Learning based SFC deployment algorithm in edge environment, where the RL agent finds the best location of each VNF, then the SFC path is obtained by connecting VNFs with shortest paths. The bandwidth constraint cannot be guaranteed in this approach. In \cite{jia2021reliability}, the authors proposed a reliability-aware RL-based SFC scheduling plan. The authors assume that all the SFC requests cannot be known in advance, and the SFC is placed by finding the location each VNF in order. In \cite{khoshkholghi2020service}, the authors proposed two heuristic algorithms for SFC placement in large scale cloud/edge environments with cost and latency considerations. In \cite{shang2019network}, the authors proposed an online SFC optimization problem and designed an approximation 
 algorithm with a theoretical performance guarantee.

\section{System Model and Problem Formulation}
\label{sec:problem_formulation}

The terms used in the system model and formulation are shown in Table \ref{table:terms}.

\subsection{Network Model}
 The edge computing network is modeled by a graph $G=(N,E)$, where $N = N_C \cup N_{AP}$ consists of a set of AP nodes, $N_{AP}$, and a set of computation nodes, $N_C$, and $E$ is a set of virtual links abstracted from the infrastructure network. Each node $n \in N_{AP}$ provides service access ingress and egress for service traffic, each node $n \in N_C$ has $C^n$ CPU cores, and each link $e \in E$ has specific bandwidth capacity $B^e$ and delay $d_T^e$. 
 
\subsection{SFC request model}
The set of SFC requests from AP $i \in N_{AP}$ is denoted as $S_i$. The $j$-th SFC request from AP $i$ is denoted as $s_{i, j}$, and is a tuple:
\begin{equation}
\label{eqn:S_i_j tuple}
\begin{aligned}
s_{i, j} = \left\{s r c, d s t, b_{i, j}, T_{i, j}^\alpha, T_{i, j}^\beta, V_{i, j}, f_{i, j}^{rep}, f_{i, j}^{aug}\right\}. \\
\end{aligned}
\end{equation}

The elements in the tuple correspond to the source AP, destination AP, required bandwidth, initiation time, expiration time, VNF list, replica vector, and boost vector. The VNF list contains the VNF instances and their execution order as follows.

\begin{equation}
\label{eqn:vnf_list}
\begin{aligned}
V_{i, j}=\left[v_{i, j}^1, v_{i, j}^2, \ldots, v_{i, j}^{\left|V_{i, j}\right|}\right]. \\
\end{aligned}
\end{equation}

Each VNF, $v_{i, j}^k$, requires a base number of CPU cores, $c_{i, j}^k$. The replica vector denotes which VNF instances require additional replicas. For example, $f_{i, j}^{rep}=[0, 1, 0]$ indicates that the second VNF in the SFC requires additional replicas to achieve higher reliability. We assume each replica of the original VNF instance requires one CPU core. Similarly, the boost vector indicates which VNF instances can be augmented with additional CPU cores beyond its base number of CPU cores to reduce processing time. For example, $f_{i, j}^{aug}=[1, 0, 1]$ indicates that the first and third VNF can reduce their processing time with additional CPU cores. The number of additional CPU cores allocated to VNF $v_{i, j}^k$ for boosting is denoted as $\sigma_{i, j}^k$.

When a SFC request is initiated, the edge operator needs to decide whether to accept this request. If yes, the SFC routing path and the VNF placement plan will be generated. Otherwise, the request will be discarded.  Then the VNF instances are deployed and activated on the selected computation nodes according to the plan, and the bandwidth of virtual links along the SFC path is also reserved. During the lifetime of a SFC, application traffic will arrive at the source AP node, receive service from the VNFs along the SFC path, and depart the edge computing network at a destination AP node. When the SFC lifetime expires, all the computation and network resources allocated to the SFC will be freed. All admission and placement decisions of SFC requests are assumed to be made by a centralized decision agent.

\begin{table}
\centering
\caption{Notation}
\label{table:terms}
\begin{tabular}{cl}
\hline Terms & Explanation \\
\hline
$S_i$ & The set of SFC requests from $i \in N_{AP}$  \\
$s_{i, j}$ & The $j$-th SFC request in $S_i$ \\
$T^{\alpha}_{i, j}$ & The arrival time of $s_{i, j}$ \\
$T^{\beta}_{i, j}$ & The departure time of $s_{i, j}$ \\
$T^{dur}_{i, j}$ & The duration time of $s_{i, j}$, $T^{dur}_{i, j}=T^{\beta}_{i, j}-T^{\alpha}_{i, j}$ \\
$E_{i, j}$ & The subset of links used by $s_{i, j}$, $E_{i, j} \subseteq E$ \\
$P_{i, j}$ & The profit of serving $s_{i, j}$ \\
$V_{i, j}$ & The set of VNFs in $s_{i, j}$ \\
$v^k_{i, j}$ & The $k$-th VNF of  $s_{i, j}$ \\
$c^k_{i, j}$ & The base number of CPU cores required by $v^k_{i, j}$ \\
$\sigma^k_{i, j}$ & The additional CPU cores allocated to $v^k_{i, j}$ for boosting \\
$r^k_{i, j}$ & The redundancy CPU cores allocated to replicas of $v^k_{i, j}$ \\
$w^k_{i, j}$ & The computation load of $v^k_{i, j}$ in cycles \\
$\tau$ & The processing speed of one CPU core in cycles per second \\
\hline
\end{tabular}
\end{table}

\subsection{Problem Formulation}
Given a set of SFC requests, the admission and placement decisions should be made. The Boolean decision variables used are given as follows:
\begin{equation}
\label{eqn:decision_x}
\begin{aligned}
x_{i, j}=\left\{\begin{array}{cc}
1, & \text { if } s_{i, j} \text { is accepted } \\
0, & \text { otherwise. }
\end{array}\right. \\
\end{aligned}
\end{equation}

\begin{equation}
\label{eqn:decision_y}
\begin{aligned}
y_{i, j}^{k, n}=\left\{\begin{array}{cc}
1, & \text { if } v_{i, j}^k \text { is deployed on node } n \in N_C\\
0, & \text { otherwise. }
\end{array}\right. \\
\end{aligned}
\end{equation}

\begin{equation}
\label{eqn:decision_z}
\begin{aligned}
z_{i, j}^e=\left\{\begin{array}{cc}
1, & \text { if } e\in E \text { is used by }s_{i, j} \\
0, & \text { otherwise. }
\end{array}\right. \\
\end{aligned}
\end{equation}

Here the variable in (\ref{eqn:decision_x}) decides whether to accept this request. If yes, then the variables in (\ref{eqn:decision_y}) and (\ref{eqn:decision_z}) decide the VNF placement locations and path selection, respectively. These decision variables need to guarantee that the following constraints will be met during the operating lifetime of the edge computing network.

The bandwidth constraint of links is denoted by (\ref{eqn:func_bw}) and (\ref{eqn:constraint_bw}). The existence function (\ref{eqn:func_bw}) requires the bandwidth can only be occupied during the lifetime of the SFC. (\ref{eqn:constraint_bw}) requires that the total bandwidth workload of all links cannot exceed their capacities at any time.

\begin{equation}
\label{eqn:func_bw}
\begin{aligned}
 b_{i, j}(t)=\left\{\begin{array}{cc}
b_{i, j}, & \text { if } t \in\left[T_{i, j}^\alpha, T_{i, j}^\beta\right] \\
0, & \text { otherwise. }
\end{array}\right. \\
\end{aligned}
\end{equation}

\begin{equation}
\label{eqn:constraint_bw}
\begin{aligned}
\sum_{i=1}^{\left|N_{AP}\right|} \sum_{j=1}^{\left|S_i\right|} x_{i, j}z_{i, j}^e b_{i, j}(t) \leq B^e, \forall e \in E.\\
\end{aligned}
\end{equation}

The computing capacity constraint of each computing node is denoted by (\ref{eqn:func_compute_dp}) - (\ref{eqn:constraint_compute}). (\ref{eqn:func_compute_dp}) requires that the CPU cores of the original VNF instance (including the additional CPU cores for boosting) and the CPU cores of its replicas should reside on the same computing node. The existence function (\ref{eqn:func_compute_time}) specifies that the CPU cores can only be occupied during the lifetime of the SFC. (\ref{eqn:constraint_compute}) indicates that the total CPU cores consumed by all VNFs in the node cannot exceed the total number of CPU cores installed on this node at any time.

\begin{equation}
\label{eqn:func_compute_dp}
\begin{aligned}
c_{i, j}^{k, n}=y_{i, j}^{k, n}\left(c_{i, j}^k+\sigma^k_{i, j}+r_{i, j}^k\right).
\end{aligned}
\end{equation}

\begin{equation}
\label{eqn:func_compute_time}
\begin{aligned}
 c_{i, j}^n(t)=\left\{\begin{array}{cl}
\sum_{k=1}^{\left|V_{i, j}\right|} c_{i, j}^{k, n}, & \text { if } t \in\left[T_{i, j}^\alpha, T_{i, j}^\beta\right] \\
0, & \text { otherwise. }
\end{array}\right. \\
\end{aligned}
\end{equation}

\begin{equation}
\label{eqn:constraint_compute}
\begin{aligned}
\sum_{i=1}^{\left|N_{A P}\right|} \sum_{j=1}^{\left|S_i\right|} x_{i, j} c_{i, j}^n(t) \leq C^n, \forall n \in N_C. \\
\end{aligned}
\end{equation}

Given a reliability bound $\Theta_{i, j}$ of $s_{i, j}$, suppose the reliability of a VNF instance is $\theta$, and let $r_{i, j}^k$ denote the number of replicas of the $k$-th VNF. Then the reliability constraint of this SFC is denoted as follows:

\begin{equation}
\label{eqn:constraint_reliability}
\begin{aligned}
\prod_{k=1}^{\left|V_{i, j}\right|}\left[1-\left(1-\theta\right)^{1+r_{i, j}^k}\right] \geq \Theta_{i, j}.
\end{aligned}
\end{equation}

The end-to-end delay constraint of a SFC comprises transmission delay and processing delay. Given a total delay upper bound of $\Phi_{i, j}$, assume the SFC has $\left|V_{i, j}\right|$ VNFs and $\left|E_{i, j}\right|$ links, then the delay constraint is denoted as follows:

\begin{equation}
\label{eqn:constraint_delay}
\begin{aligned}
\sum_{e=1}^{\left|E_{i, j}\right|} d_T^e+\sum_{k=1}^{\left|V_{i, j}\right|} d_p^k \leq \Phi_{i, j}.
\end{aligned}
\end{equation}

Here $d_T^e$ denotes the transmission delay on link $e$, $d_p^k$ denotes the processing delay of $v_{i, j}^k$, which can be further denoted as follows:

\begin{equation}
\label{eqn:process_delay}
\begin{aligned}
d_p^k=\frac{w_{i, j}^k}{(c_{i, j}^k+\sigma^k_{i, j}) \tau}.
\end{aligned}
\end{equation}

The goal of this problem is to maximize the profit of the edge service provider, where the objective function is given as follows:

\begin{equation}
\label{eqn:objective}
\begin{aligned}
& \max P=\sum_{i=1}^{\left|N_{AP}\right|} \sum_{j=1}^{\left|S_i\right|} P_{i, j}=\sum_{i=1}^{\left|N_{AP}\right|} \sum_{j=1}^{\left|S_i\right|} x_{i, j} b_{i, j} C_{i, j} T_{i, j}^{dur} \eta_{i, j} \\
& \text { s.t. (7), (10) - (12). } \\
&
\end{aligned}
\end{equation}

Here $C_{i, j}=\sum_{k=1}^{\left|V_{i, j}\right|}c_{i, j}^k$ is the total CPU consumption of $s_{i, j}$, and $\eta_{i, j}$ is the penalty factor for extra CPU consumption made by boosting and replica allocation.

\begin{equation}
\label{eqn:penalty_factor}
\begin{aligned}
\eta_{i, j}=\frac{C_{i, j}}{C_{i, j}+\sum_{k=1}^{\left|V_{i, j}\right|}r_{i, j}^k+\sigma_{i, j}^k}\leq1.
\end{aligned}
\end{equation}

As the formulation shows, the delay-aware and reliability-aware SFC placement problem is a Mixed Integer Non-Linear Problem (MINLP) which is hard and intractable. In \cite{6867768}, the SFC placement problem is reduced to Flexible Job-shop Scheduling Problem, which is NP-hard. Thus, our problem is NP-hard as well. In next section, we present our multi-agent RL-based SFC placement scheme to obtain a sub-optimal solution.

\section{Proposed Scheme}
\label{sec:Proposed_Scheme}

\subsection{Scheme Overview}
In our scheme, we apply a Reinforcement Learning (RL) framework to the SFC placement problem and develop concrete implementations of the environment and agent. Each interaction loop represents the decisions for one SFC request. The environment is comprised of the edge computing network and the current SFC request. The state observation information includes the current network state and the SFC request features. The agent acts as the network operator and is implemented as a multi-agent system, where different agents collaborate in a cascading order and generate a collection of decisions. First, the path agent makes an admission and path selection decision. With the path selection result and the SFC request features, the VNF configuration agent allocates concrete CPU cores for each VNF of the SFC. Finally, the pattern agent generates a pattern decision, also known as the VNF placement plan. The action comprises all the decisions made by the multi-agent system. The edge network deploys the SFC according to the action and then updates the current network state. The reward for serving this SFC is generated and sent to the multi-agent system. The multi-agent system updates the policy on each agent based on the reward.

For the path agent and the pattern agent, we adopt Deep Neural Networks (DNNs). For the VNF configuration agent, we utilize a heuristic algorithm, since this agent only needs to determine the number of CPU cores allocated to each VNF while satifying all the constraints. In Section \ref{sec:Proposed_Scheme}-B and \ref{sec:Proposed_Scheme}-C, we present the detailed design of the environment and the multi-agent system. In Section \ref{sec:Proposed_Scheme}-C, we describe the RL algorithm and our adaptation.

\subsection{Environment Design}
To represent the observation of network state and the SFC request as the input of the DNN based agent, we encode them as specific vectors and concatenate them together as the observation vector. The network state is encoded as a residual resources vector, in which the elements of the vector denote the residual number of CPU cores at each computing node and the residual amount of bandwidth on each link. The size of this vector is $\left|N\right|+\left|E\right|$. The features of the SFC request are also encoded as a feature vector. To denote the source and destination, we employ one-hot encoding using a vector of size $\left|N\right|$, where each element corresponds to one node in the topology. The elements corresponding to the source node and destination node are set to 1, and the other elements are set to 0. Each scalar value of the request, such as bandwidth and duration time, is encoded by a single element. For the CPU requirement of VNFs, we use a vector of size $\left|V_{i,j}\right|$ and mark each element with the number of CPU cores required by each VNF. The feature vectors for the number of replicas and the number of additional CPU cores for boosting have a size of $\left|V_{i,j}\right|$, and their format is consistent with the definitions in Section \ref{sec:problem_formulation}-B. By concatenating all the vectors together, we obtain an observation vector with size $2\left|N\right|+\left|E\right|+ 3+ 3\left|V_{i,j}\right|$.

\begin{figure}[tb]
\centerline{\includegraphics[width=0.35\textwidth]{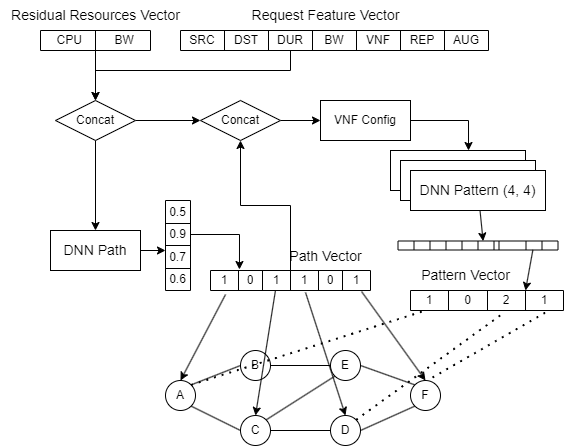}}
\caption{Workflow of Multi-Agent System}
\label{fig:agents}
\end{figure}

\subsection{Agent Design}
The multi-agent system is comprised of a path agent, a VNF configuration agent, and a group of pattern agents. Here we describe their collaboration process and depict it as a workflow shown in Fig. \ref{fig:agents}.

First, the state observation vector serves as the input of the DNN for path selection (this DNN is named as DNN-path). The output of DNN-path is a fixed size vector. In this vector, apart from the first element, which indicates acceptance or rejection, each element denotes the probability of choosing a corresponding candidate path. For example, if there are 3 candidate paths for selection, then the size of the output vector of DNN-path is 4. The path corresponding to the element with maximal value will be the path selection result. The resulting path is then encoded using one-hot encoding and this vector serves as the output vector of the path agent. In Fig. \ref{fig:agents}, suppose the resulting path is ``A-C-D-F". Then the output vector will be [1,0,1,1,0,1].
The path vector is then concatenated with state observation vector to form the pattern observation vector, which acts as the input to the VNF configuration agent. 

The VNF configuration agent will adjust the number of CPU cores for each VNF to meet the constraints in Section \ref{sec:problem_formulation}. The heuristic algorithm simply increases the number of CPU cores of the first feasible VNF (obtained from the $f_{i, j}^{rep}, f_{i, j}^{aug}$ fields in the request feature vector) by 1 to see if the constraints hold. If not, the process is repeated for the next feasible VNF. If no solution is found, then the request is dropped. If a solution exists, then the updated pattern observation vector serves as the input to the pattern agent.

Assume that the path obtained from the path agent has $m$ computing nodes, and the SFC request requires $n$ VNFs. Then the number of possible deployment patterns can be denoted as follows:

\begin{equation}
\label{eqn:pattern_num}
\begin{aligned}
P(n, m)=\sum_{i=1}^{\min (m, n)}\left(\begin{array}{c}
m \\
i
\end{array}\right)\left(\begin{array}{c}
n-1 \\
i-1
\end{array}\right).
\end{aligned}
\end{equation}

The first combination selects $i$ nodes out of $m$ on which to deploy VNFs, (each selected node needs to deploy at least one VNF), and the second combination indicates the feasible number of allocation arrangements of $n$ VNFs to $i$ nodes. For example, $P(3, 2) = 4$, $P(2, 3) = 6$ and $P(3, 3)=10$. As $m$ and $n$ increase, $P(m, n)$ grows rapidly. To control the space complexity, we limit the range of $m$ and $n$, i.e., $m,n \in [2, 4] $. For each combination of $m$ and $n$, the input and output vector size is fixed, and a dedicated DNN (this DNN is named as DNN-pattern) is assigned to make the pattern selection. Namely, $3\times3=9 $ DNNs serve as 9 pattern agents in our scheme. The output of DNN-pattern is a vector of size $P(m, n)$. In this vector, each element denotes the probability of choosing the corresponding pattern. The pattern corresponding to the element with maximal value will be the pattern selection result. The vector notation of the resulting pattern serves as the output of this pattern agent. In Fig. \ref{fig:agents}, the SFC request needs to deploy 4 VNFs on path ``A-C-D-F", the DNN-pattern agent for P(4, 4) will handle this request and output the pattern vector [1, 0, 2, 1], which indicates that the 1st VNF is deployed on node A, the 2nd and 3rd VNFs are deployed on node D, and the 4th VNF is deployed on node F. 

\subsection{Training Algorithm}

The RL algorithm adopted here is Deep Q Network (DQN) \cite{mnih2013playing}, an off-policy and model free algorithm that learns via Q-learning. In Q-learning, the state $s$ and action $a$ serve as input, and a reward is obtained as output. The goal is to find the optimal policy $\pi$  to obtain the optimal Q-value function $Q^*(s, a)=\max _\pi \mathbb{E}\left[\sum_{k=0}^{\infty} \gamma^k r_{t+k} \mid s_t=s, a_t=a, \pi\right]$, where the future rewards are discounted by $\gamma$ per time step. After each action, a state transition experience $e_t=\left(s_t, a_t, r_t, s_{t+1}\right)$ can be obtained, and the Q-value can be updated as $Q\left(s_t, a_t\right) \leftarrow Q\left(s_t, a_t\right)+\alpha\left(y_t-Q\left(s_t, a_t\right)\right)$, in which $y_t=$ $r_{t+1}+\gamma \max _{a_{t+1}} Q\left(s_{t+1}, a_{t+1}\right)$, and $\alpha$ is the learning rate. $y_t$ is the target and should be equal to $Q\left(s_t, a_t\right)$ if the learning process converges. DQN does not calculate Q-values explicitly, but uses a DNN to approximate the optimum Q-values: $Q^*(s, a) \approx$ $Q(s, a ; \theta)$. $\theta$ is the weight parameter function of the DNN.

To stabilize the learning performance, DQN introduces replay memory pooling, Eval-Net and Target-Net. Replay memory pool buffers the transition experiences and randomly takes samples from it for the learning. This random sampling prevents DQN from undergoing fluctuation due to learning from correlated experiences in sequences. The learning process of DQN updates not only $Q\left(s_t, a_t\right)$ but also the target value $y_t$, since $y_t$ involves the estimate of the Q-value. The Eval-Net learns for each experience, and the Target-Net calculates the target value $y_t$. The weight function $\theta_T$ of Target-Net is copied from the weight function $\theta$ of Eval-Net periodically.

In our multi-agent RL scheme, the path agent and the pattern agents use DQN to improve their decision performance. They maintain their own replay memory pools and update their own DNN parameters asynchronously. For each SFC request placement process, the path agent and one of the pattern agents that serves this request can accumulate one transition experience.

\section{Experiment and Evaluation}
\label{sec:experiment}

\subsection{Experiment Settings}

\begin{figure}[tb]
\centerline{\includegraphics[width=0.35\textwidth]{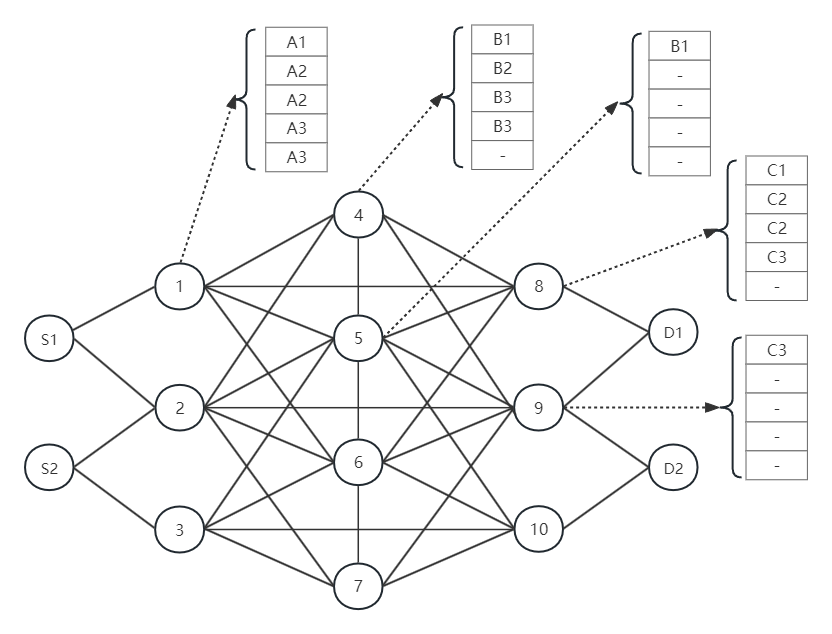}}
\caption{Experiment Topology}
\label{fig:topo}
\end{figure}

The topology of edge computing networks used in simulation is shown in Fig. \ref{fig:topo}, and it contains two source AP nodes, two destination AP nodes, and ten computation nodes. Each computation node is equipped with 32 CPU cores. The bandwidth capacity allocated to each link in the topology is randomly selected from (10 GB/s, 15 GB/s, 20GB/s). The edge computing network generated from this topology operates 200 logical time slots.
The SFC requests are generated following a Poisson arrival process with parameter $\lambda$, and the service holding time follows an exponential distribution with parameter $\mu$. For the remaining features in the request tuple, the source and destination node are randomly selected from the AP nodes; the bandwidth requirement is randomly selected from (200 MB/s, 500 MB/s, 1GB/s); the basic CPU requirement of each VNF is randomly generated within the range 1 to 4; the Boolean feature vectors $f_{i, j}^{rep}, f_{i, j}^{aug}$ are also randomly generated.

\subsection{Baseline Algorithms}

Our RL-based scheme is compared with a heuristic multi-agent system which adopts two heuristic algorithms for path and pattern selection. The path heuristic algorithm filters out those paths whose residual bandwidth cannot accommodate the SFC traffic and chooses the path with maximal residual CPU cores. The pattern heuristic algorithm finds the location of each VNF in their serving order along the selected path, and deploys each VNF on the first feasible node found whose residual CPU cores meet its requirement. After all the locations of VNFs are found, a pattern is obtained.

In our simulation, the multi-agent system has four implementations. Our proposed scheme uses a RL path agent and a set of RL pattern agents and is denoted as ``RL + RL"; the pure heuristic scheme uses a heuristic path agent and a heuristic pattern agent and is denoted as ``H + H". We also evaluate the mixture implementations. The combination of a RL path agent and a heuristic pattern agent is denoted as ``RL + H", and the combination of a heuristic path agent and a set of RL pattern agents is denoted as ``H + RL".

\subsection{Reinforcement Learning Settings}
For Target-Net and Eval-Net of DQN, we implement 5 fully-connected layers, and each layer contains $256$ hidden neurons. The activation function is $tanh$. The initial learning rates is $0.001$. The discount factor $\gamma$ is set within the range of [0,1, 1] with interval of 0.1 to see its performance on total rewards. Instead of updating model parameters every step, we update parameters every 5 steps. To avoid the lack of data at the beginning of training, we collect $2000$ pieces of transition experiences for each RL agent before learning. The reward for each deployed SFC is computed following the definition in (\ref{eqn:objective}), and the reward for a rejected request is 0. We set the RL algorithm to run for 700 episodes. In each episode, the multi-agent RL system gives placement decisions for all SFC requests arriving in the logical simulation time of the edge computing network.

\begin{figure}[tb]
\centerline{\includegraphics[width=0.43\textwidth]{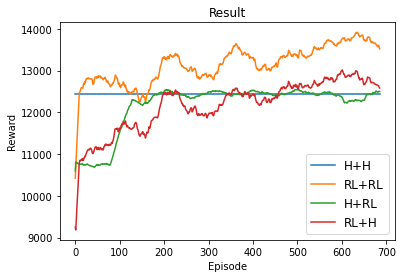}}
\caption{Performance Comparison}
\label{fig:fourline_compare}
\end{figure}

\begin{figure}[tb]
\centerline{\includegraphics[width=0.35\textwidth, height=0.25\textwidth]{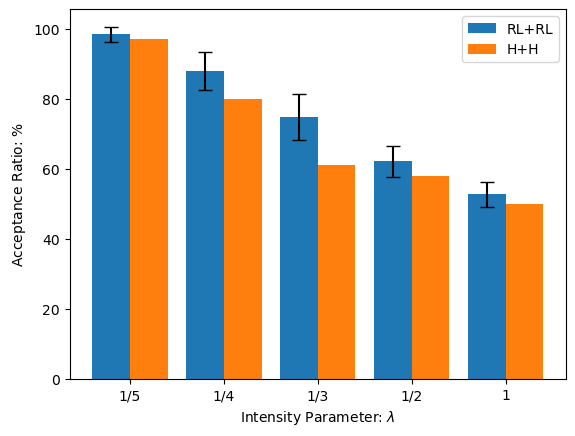}}
\caption{Influence of Arrival Intensity}
\label{fig:accept_ratio}
\end{figure}

\subsection{Performance Results}
We compare our proposed scheme with three baseline schemes as shown in Fig. \ref{fig:fourline_compare}. The performance of our multi-agent RL scheme converges when the training passes around 600 episodes, and surpasses the pure heuristic scheme ``H+H " by at least 12$\%$. The mixture implementation ``RL+H" also performs better than the pure heuristic solution, but the advantage is less than 5$\%$. The mixture implementation ``H+RL" performs almost the same as the pure heuristic solution. From the result, we conclude that the collaboration of RL agents can achieve the best performance among all implementations.

In order to explore the performance of our scheme in terms of acceptance ratio with different arrival intensities, we vary $\lambda$ as the intensity parameter. The result is depicted in Fig. \ref{fig:accept_ratio}. When $\lambda$ is close to 1, both our scheme and the heuristic solution cannot accommodate as many requests, and the acceptance ratio is around 50$\%$. When $\lambda$ is equal or smaller than 1/5, the arrival intensity is low enough and both schemes can successfully accept most requests. When the $\lambda$ equals 1/3, our scheme achieves the highest acceptance ratio advantage over the heuristic solution.

\begin{figure}[tb]
\centerline{\includegraphics[width=0.35\textwidth,height=0.25\textwidth]{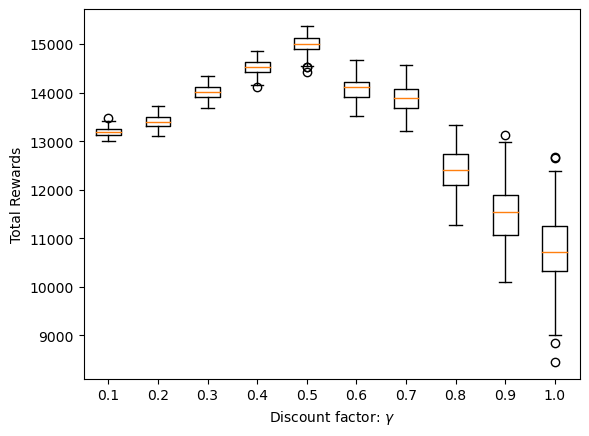}}
\caption{Performance Comparison on Discount Factor}
\label{fig:discount_gamma}
\end{figure}

We also evaluate the influence of discount factor $\gamma$ on the total reward. The result is shown in Fig. \ref{fig:discount_gamma}. When $\gamma$ equals 0, the RL algorithm will reduce to an algorithm with an exploitation strategy and focus on immediate rewards. Our scheme can achieve optimal performance when $\gamma$ equals 0.5. If $\gamma$ continues growing and approaches 1, the future reward will have dominant influence on the current decision, the variance of total reward will increase, and the performance will deteriorate. 

\section{Conclusion}
\label{sec:conclusion}
In this paper, we formulate the edge SFC placement problem with multiple constraints. We propose a multi-agent reinforcement learning scheme to address this problem with latency and reliability considerations. The agents in the proposed scheme collaborate to make SFC placement decisions, including route path selection, VNF configuration and VNF placement. The experimental results show that our scheme outperforms the heuristic solution in terms of acceptance ratio and total reward.

In our current work, we limit the computing node number $m$ of the path and the VNF number $n$ of a SFC to a small range. The purpose is to use a dedicated RL agent to train the pattern selection policy for every $(m, n)$ combination. In the future, we plan to use a grouping strategy to deal with the pattern selection problem for paths with more computing nodes and SFCs with more VNFs.  

\section{ACKNOWLEDGMENT}
This work was supported in part by the National Science Foundation under Grant No. CNS-2008856.

\bibliographystyle{ieeetr}
\bibliography{reference}

\begin{thebibliography}{1}

\bibitem{8016573}
Y.~Mao, C.~You, J.~Zhang, K.~Huang, and K.~B. Letaief, ``A survey on mobile edge computing: The communication perspective,'' {\em IEEE Communications Surveys \& Tutorials}, vol.~19, no.~4, pp.~2322--2358, 2017.

\bibitem{alleg2017delay}
A.~Alleg, T.~Ahmed, M.~Mosbah, R.~Riggio, and R.~Boutaba, ``Delay-aware vnf placement and chaining based on a flexible resource allocation approach,'' in {\em 2017 13th international conference on network and service management (CNSM)}, pp.~1--7, ieee, 2017.

\bibitem{akahoshi2021service}
K.~Akahoshi, F.~He, and E.~Oki, ``Service deployment model with virtual network function resizing,'' in {\em 2021 IEEE Global Communications Conference (GLOBECOM)}, pp.~1--6, IEEE, 2021.

\bibitem{9236981}
S.~Pandey, J.~W.-K. Hong, and J.-H. Yoo, ``Q-learning based sfc deployment on edge computing environment,'' in {\em 2020 21st Asia-Pacific Network Operations and Management Symposium (APNOMS)}, pp.~220--226, 2020.

\bibitem{jia2021reliability}
J.~Jia, L.~Yang, and J.~Cao, ``Reliability-aware dynamic service chain scheduling in 5g networks based on reinforcement learning,'' in {\em IEEE INFOCOM 2021-IEEE Conference on Computer Communications}, pp.~1--10, IEEE, 2021.

\bibitem{khoshkholghi2020service}
M.~A. Khoshkholghi, M.~Gokan~Khan, K.~Alizadeh~Noghani, J.~Taheri, D.~Bhamare, A.~Kassler, Z.~Xiang, S.~Deng, and X.~Yang, ``Service function chain placement for joint cost and latency optimization,'' {\em Mobile Networks and Applications}, vol.~25, pp.~2191--2205, 2020.

\bibitem{shang2019network}
X.~Shang, Z.~Liu, and Y.~Yang, ``Network congestion-aware online service function chain placement and load balancing,'' in {\em Proceedings of the 48th International Conference on Parallel Processing}, pp.~1--10, 2019.

\bibitem{6867768}
J.~F. Riera, E.~Escalona, J.~Batallé, E.~Grasa, and J.~A. García-Espín, ``Virtual network function scheduling: Concept and challenges,'' in {\em 2014 International Conference on Smart Communications in Network Technologies (SaCoNeT)}, pp.~1--5, 2014.

\bibitem{mnih2013playing}
V.~Mnih, K.~Kavukcuoglu, D.~Silver, A.~Graves, I.~Antonoglou, D.~Wierstra, and M.~Riedmiller, ``Playing atari with deep reinforcement learning,'' {\em arXiv preprint arXiv:1312.5602}, 2013.

\end{thebibliography}
\end{document}